\documentclass[%
    reprint,
    amsmath,amssymb,
    longbibliography
]{revtex4-1}

\usepackage{graphicx}% Include figure files
\usepackage{dcolumn}% Align table columns on decimal point
\usepackage{bm}% bold math
\usepackage{amsmath}
\usepackage[utf8]{inputenc}
\usepackage{caption}
\usepackage{makecell}
\usepackage{xcolor}
\usepackage{textgreek}
\usepackage{tabularx}
\usepackage{graphicx} % 在导言区添加
\usepackage{diagbox}
\usepackage{etoolbox}

\usepackage{multirow}

\usepackage{array}      % 自定义列格式

\begin{document}
\captionsetup{justification=raggedright, singlelinecheck=false}
\preprint{APS/123-QED}

\title{Self-assembling clusters of particles on a shrinking liquid surface}% Force line breaks with \\

\author{Xin Li,$^1$ Shuchen Zhang,$^1$ Mark J. Bowick,$^{2,}$}
\email{bowick@kitp.ucsb.edu}
\author{Duanduan Wan$^{1,}$}
\email{ddwan@whu.edu.cn}

\affiliation{$^1$Key Laboratory of Artificial Micro- and Nano-structures of Ministry of Education and School of Physics and Technology, Wuhan University, Wuhan 430072, China \\
$^2$Kavli Institute for Theoretical Physics, University of California, Santa Barbara, California 93106, USA}

\date{\today}% It is always \today, today,

\begin{abstract}
After rainfall, pine needles often float on the surface of small puddles. As the water evaporates, they self-assemble into distinct clusters. Motivated by this natural phenomenon, we experimentally investigate the dynamic evolution of synthetic particles as the liquid surface shrinks in area. 
Our experiments demonstrate the tendency of particles to aggregate, forming distinct clusters as the liquid boundary shrinks. We systematically examine the emergence of these clusters and explore how their sizes and numbers evolve with changes in packing fraction. We also analyze particle rotation during the process and discuss the formation of the final configuration comprising clusters of various orientations. Complementary numerical simulations demonstrate qualitative agreement with our experimental findings. This study sheds additional light on the self-assembly of macroscopic particles in a dynamically evolving medium. 
\end{abstract}

\maketitle

%\tableofcontents

\section{\label{sec:level1}INTRODUCTION}

It is of significant interest to understand the ability of particles to self-assemble into complex structures. By exploring how this collective behavior emerges, we can gain insight into fundamental questions of how matter organizes itself, essential to condensed matter physics, materials science, and even our understanding of life itself \cite{Manoharan2015}. Self-assembly of particles is mostly observed in colloidal systems (e.g., Refs.~\cite{Michael2016Self, Li2016AssemblyAP, Li2021RevealingTN, Peng2023}). Recent advancements in synthesis have led to the production of a diverse array of anisotropic particles, which readily organize into various superlattices under the appropriate experimental conditions (e.g., Refs.~\cite{Henzie2012, Young2013, Gong2017, Forster2011, Hosein2010, Meijer2017, Miszta2011}). When the predominant interparticle interaction is the hard core potential of particle shape and other interactions are weak, colloidal particles can be approximated as hard particles. Simulations of hard colloids (e.g., Refs.~\cite{Torquato2009,  Marechal2010, Gang2011, Agarwal2011,  Smallenburg2012, Avendano2012, Gantapara2013, Bernard2011, Wan2018, Wan2022_PRE, Wan2023_SoftMatter}) predict complex crystals from an even larger variety of anisotropic shapes.

As particles increase in size, they become less susceptible to motion induced by thermal fluctuations of the solvent molecules, limiting self-assembly in everyday scenarios. Capillary forces at fluid interfaces, however, enable self-assembly across a wide range of particle sizes, from the microscopic to the macroscopic~\cite{ Nature2000.6790.1033, Eur.Phys.J.E.2013.36.127}. For particles floating on a fluid surface, the interface deforms due to a combination of buoyancy and surface tension forces, creating a meniscus around each particle~\cite{Langmuir2010.19.15142}. Depending on the meniscus shape, which is influenced by factors such as surface tension, particle weight,  wetting conditions, and particle shape, either attractive or repulsive interactions can occur~\cite{PNAS2011.52.20923, PNAS2017.39.10373}. When particles are identical, they often exhibit capillary attraction (e.g., Refs.~\cite{Nicolson1949, Mahadevan2005, softmatter2009, Vella2015_review, Protiere2023, Bowden1997}), a phenomenon exemplified by the “Cheerios effect", where pieces of breakfast cereal floating on the surface of a bowl tend to cluster together.

After rainfall it is frequently observed that fallen pine needles in a small puddle first float on the water surface. As water evaporates from the puddle, however, pine needles spontaneously form distinct clusters of varying orientations [see Fig.~\ref{fig_setup}(a)]. The dashed blue line in the figure indicates the outline of a cluster. In this cluster, pine needles have a similar orientation indicated by the blue arrow. Here we ask the question: how do these clusters of various orientations form and settle into well-defined structures?

We attribute this clustering phenomenon to two main factors: the Cheerios effect, which brings the needles together, and the evaporation of water, leading to a reduction in surface area and a consequent increase in packing density. To systematically investigate this process, wherein macroscopic particles at the liquid surface self-assemble while the surface area reduces, we conducted experimental studies [see Fig.~\ref{fig_setup}(b)]. We characterized the emergence of clusters, examined the evolution of cluster sizes and their number as a function of packing fraction, and analyzed particle rotation during the process.  Additionally, we conducted numerical simulations that account for the adhesion of particles and the reduction of surface area, with results qualitatively agreeing with our experimental findings. We discuss the formation of the final configuration comprising clusters of various orientations.

\section{\label{sec:level1}Experimental Methods}

\noindent {\normalsize \textbf{Preparation of particles}}
\vspace{0.5em}

\noindent White polylactic acid (PLA) particles were fabricated using a Flashforge Guider II S 3D printer. Each particle measures 35 mm in length, 7 mm in width, and 7 mm in height, with a centrally located groove (20 mm × 4 mm × 1 mm) on each of its four major faces. The particles were designed to be partially filled, achieving a density of 0.75 $\text{g}/\text{cm}^3$, which enables them to float on the oil surface (see Appendix \ref{appendix_particle_structure} for details on their structure and floating behavior). To facilitate position and orientation tracking via image analysis, the grooves on all four major faces were marked with red indicators.  

\vspace{1em}
\noindent {\normalsize \textbf{Experimental setup}}
\vspace{0.5em}

\begin{figure}[htp]
\centering
\includegraphics[width=.9\linewidth]{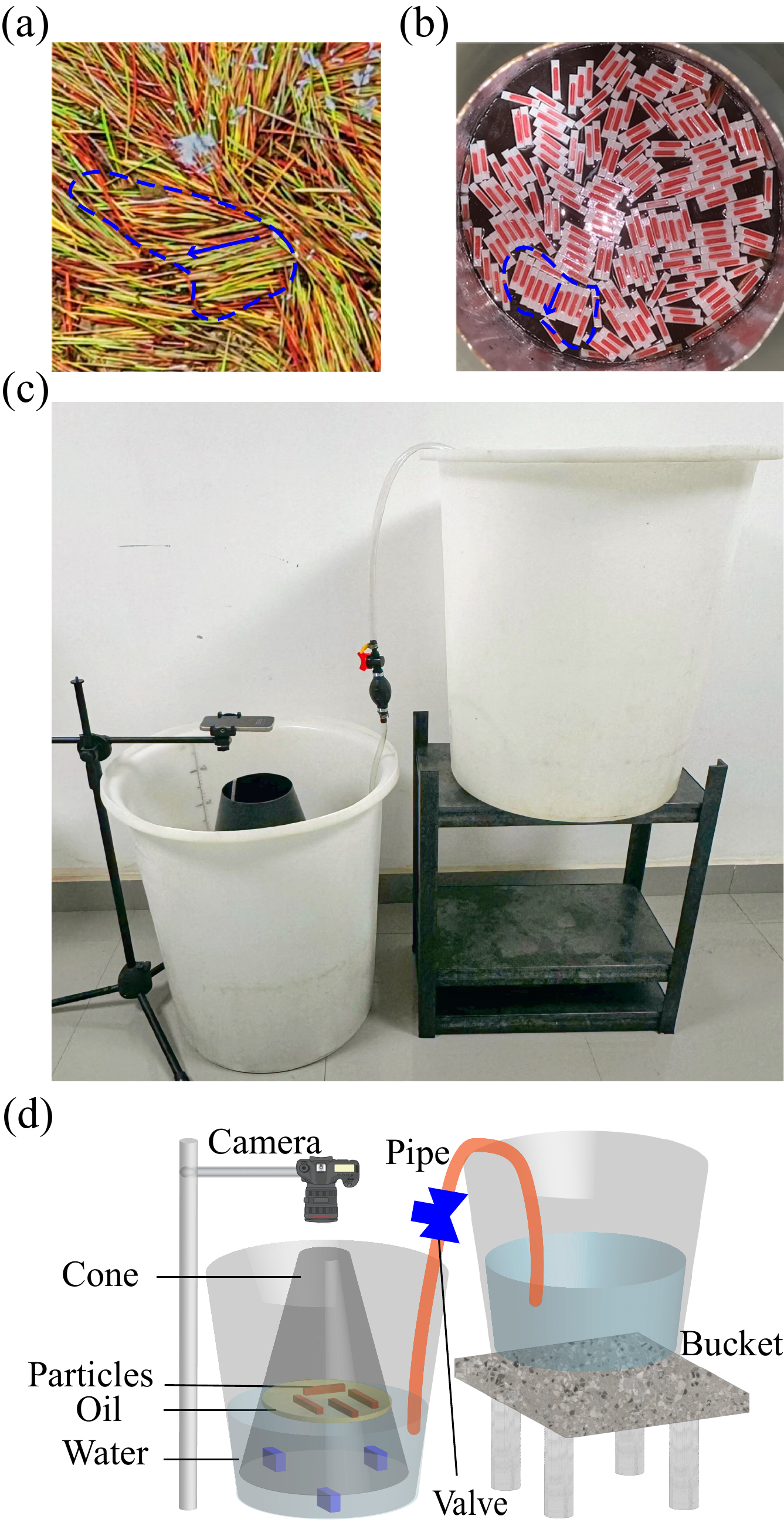}
\caption{Illustration and experimental setup. (a) The spontaneous clustering of pine needles in a puddle after rain, where clusters are defined by strong orientational alignment. The dashed blue line outlines a given cluster, with the blue arrow showing the average orientation of that cluster. (b) A snapshot, from experiment, at high density packing density. As in (a), the dashed blue line marks a cluster and the blue arrow is the associated orientation. (c) Photograph of the experimental setup. (d) Schematic of the experimental setup.}
\label{fig_setup}
\end{figure}

\noindent Figure~\ref{fig_setup}(c) shows the experimental setup which consists of two identical buckets arranged vertically. The lower bucket rests on the floor while the higher one sits atop a shelf to generate a vertical gradient. The elevated bucket serves as a water-filled reservoir.
The two buckets are connected by a pipe equipped with a valve and so water siphons from the higher bucket to the lower one. In the lower bucket a metal cone in the form of a frustum is strategically positioned on three small blocks, elevating it above the bucket's bottom [see Fig.~\ref{fig_setup}(d) for an illustration of the experimental setup]. This setup enables water to enter from the base, connecting the cone with the bucket. Initially, we fill the lower bucket with water until it submerges the three supporting blocks of the metal cone. To confine particles on the liquid surface in the subsequent step, we add a layer of corn oil to the water surface. The oil, with a density of 0.91 g/cm³, forms a layer approximately 1.57 cm thick. Subsequently, we place 150 particles sparsely across the oil surface. A camera is positioned overhead to record the experiments. The initial diameter of the oil surface within the cone measures approximately 44 cm, resulting in an initial packing density of about 0.24 [Fig.~\ref{fig_self_assembly}(a)]. The experiment starts with opening the valve, enabling water to flow from the higher bucket to the lower one at the slowest possible rate. As the surface level rises, its area decreases due to the cone's shape.

\vspace{1em}
\noindent {\normalsize \textbf{Image analysis}}
\vspace{0.5em}

\noindent We utilized OpenCV (Open Source Computer Vision Library, implemented in Python) for particle identification. This software enables the detection of the minimum rectangle encapsulating the red rectangular marker within the field of view. After obtaining the coordinates of the four vertices of the red markers, we ascertain both the center positions and orientations of the particles.

\section{\label{sec:level1}Results and Discussion}
\begin{figure*}[htp]
\centering
\includegraphics[width=17.8cm]{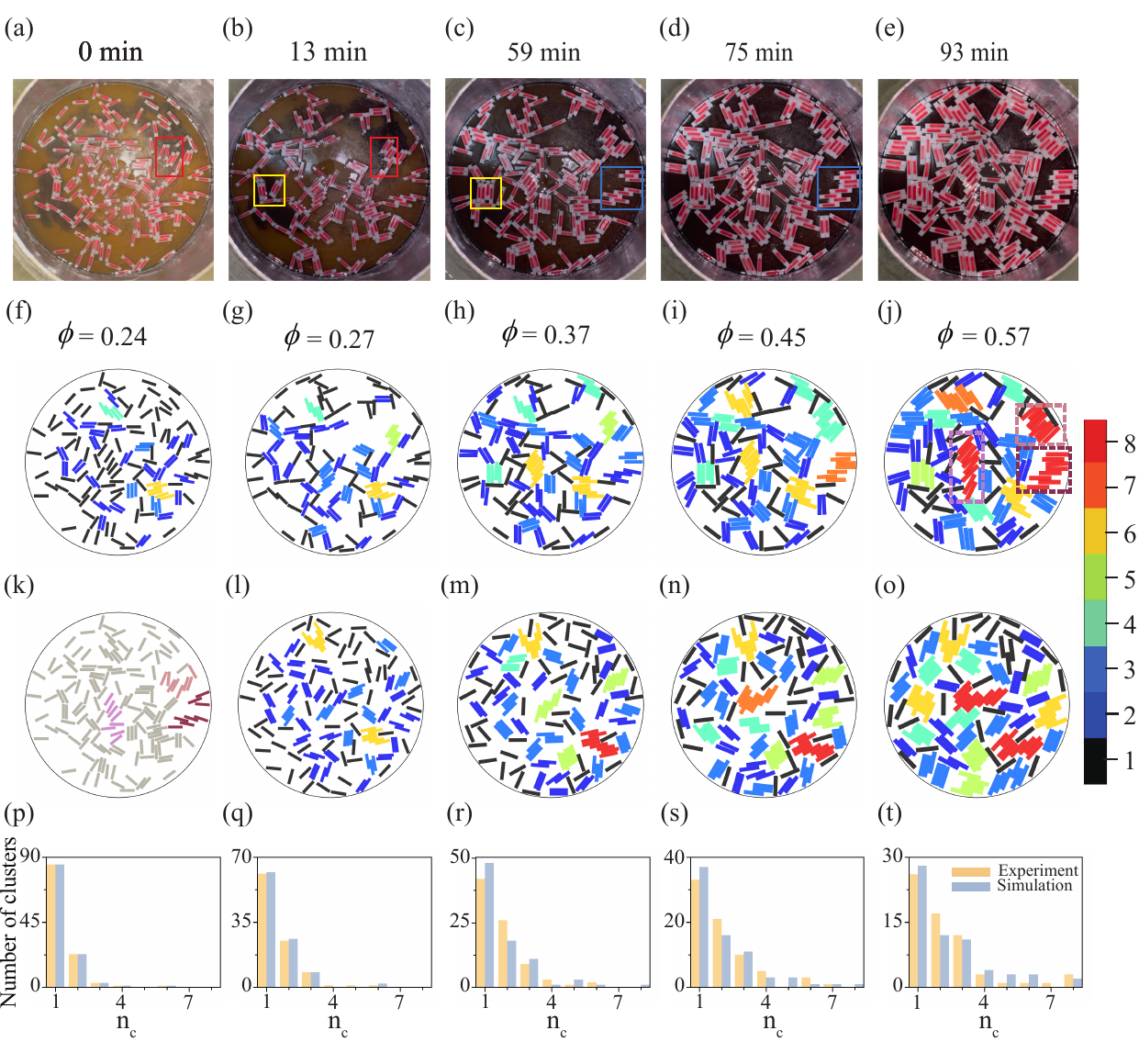}
\caption{Formation of clusters. (a)-(e) Sequential photographs of cluster evolution at 0, 13, 59, 75, and 93 minutes, respectively. Nearby particles, or clusters, merge into larger structures highlighted by red, yellow, and blue windows. (f)-(j) Visualization of cluster identification, with cluster colors representing respective sizes (number of particles). The maroon, palevioletred, and orchid windows in (j) denote the three largest clusters, each containing 8 particles. (k)-(o) Corresponding configurations in simulations at matching packing densities to (f)-(j). In (k), maroon, palevioletred, and orchid particles represent those forming the largest clusters in (j), while the others are depicted in grey. (p)-(t) Distribution of cluster size $n_{c}$, corresponding to the configurations in (f)-(j), experiment, and (k)-(o) simulation.}
\label{fig_self_assembly}
\end{figure*}

Figures \ref{fig_self_assembly}(a)-\ref{fig_self_assembly}(e) depict the evolution of particle configurations as the surface level rises. The packing density collectively increases as the surface area decreases, with neighboring particles adhering to form clusters. We highlight several examples using red, yellow, and blue windows in the figures. To gain deeper insight into cluster formation, we analyze experimental configurations and identify clusters within the configurations (see Appendix \ref{appendix_identification_clusters} for cluster identification). Clusters are color-coded based on their size, with particles in the same cluster sharing the same color, as illustrated in Figs.~\ref{fig_self_assembly}(f)-\ref{fig_self_assembly}(j). Initially, the configuration displays only one largest cluster of size 6 [highlighted in yellow in Fig.~\ref{fig_self_assembly}(f)]. As the experiment progresses, larger clusters emerge. The experiment lasts for approximately 93 minutes. The final configuration reaches a packing density of approximately 57\%, featuring three large clusters of size 8 [Fig.~\ref{fig_self_assembly}(j)]. These three largest clusters are marked in red and emphasized by maroon, palevioletred, and orchid windows, respectively.

\begin{figure}[htp]
\centering
\includegraphics[width=0.9\linewidth]{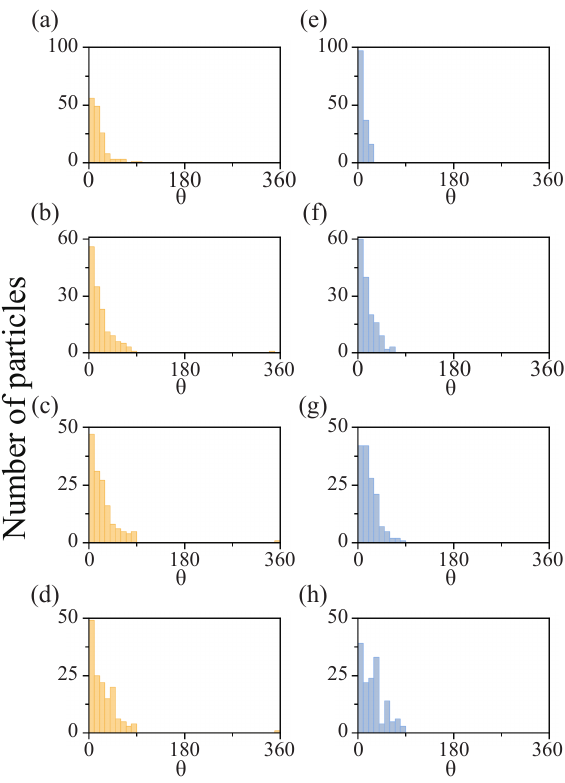}
\caption{Particle rotation during the process. (a)-(d) Distribution of rotation angle $\theta$ obtained from experimental data, corresponding to the configurations illustrated in Fig.~\ref{fig_self_assembly}(b)-\ref{fig_self_assembly}(e), respectively. Each bin represents a 10-degree width.(e)-(h) Distribution of rotation angle $\theta$, extracted from simulations conducted at identical packing densities to those in (a)-(d), corresponding to the configurations depicted in Fig.~\ref{fig_self_assembly}(l)-\ref{fig_self_assembly}(o), respectively.}
\label{fig_rotation_angle}
\end{figure}

\begin{figure}[htp]
\centering
\includegraphics[width=1\linewidth]{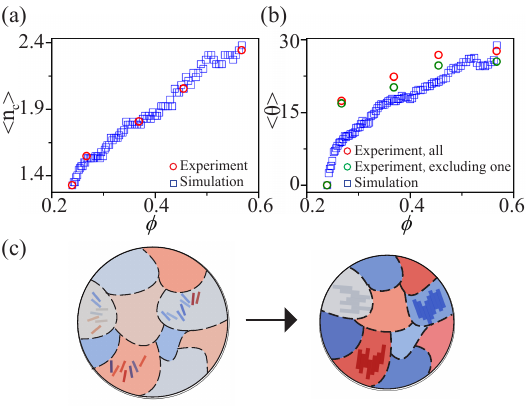}
\caption{Cluster formation with various orientations. (a) Average cluster size $<n_{c}>$ as a function of packing density $\phi$. (b) Average rotation angle $<\theta>$ as a function of packing density $\phi$. (c) Illustration of cluster formation with various orientations. Colors indicate orientation.}
\label{fig_average_size}
\end{figure}

We next perform Monte Carlo simulations of the self-assembly process. Here, we model particles as two-dimensional hard rectangles and employ hard particle Monte Carlo simulations (e.g., Refs.~\cite{Haji-Akbari2009, Damasceno2012, Ni2012, Anderson2017, Wan2021, Wan2023_PRB, Lei2018_helix, Klotsa2018, Wan2019}) augmented with an adhesion mechanism between particles inspired by the Cheerios effect. When the outermost particles of two adjacent clusters (or individual particles) are within an interaction range dictated by the Cheerios effect \cite{Mahadevan2005}, and they are similarly oriented, we allow them to adhere with a specified probability. For simplicity, we only consider the attraction between the two outermost particles and focus solely on the interaction between their elongated sides, neglecting influences from other particles or clusters. The final overlap length of the particles, as a function of the initial overlap length, is determined from experiments on two-particle attraction (see Appendix \ref{appendix_two_particle_attraction} for details). To accommodate for area shrinkage the circular boundary was gradually contracted in the simulations. We execute the hard particle Monte Carlo simulations with the adhesion mechanisms until finding configurations that accommodate the boundary contraction. These configurations are recorded at specific packing densities. Detailed simulation procedures are provided in Appendix \ref{appendix_simulation_details}.

To align closely with the experimental setup,  we initialized the simulation with the same configuration as in the experiment, depicted in Fig.~\ref{fig_self_assembly}(k), mirroring Fig.~\ref{fig_self_assembly}(f). In Fig.~\ref{fig_self_assembly}(k), we illustrate particles forming the three largest clusters in the final experimental configuration [Fig.~\ref{fig_self_assembly}(j)] in maroon, palevioletred, and orchid, respectively, reflecting the color scheme of the corresponding windows in Fig.~\ref{fig_self_assembly}(j), while other particles are depicted in grey. Notably, particles forming the same clusters in the final configuration originate from close proximity in the initial configuration. Figures ~\ref{fig_self_assembly}(l)-\ref{fig_self_assembly}(o) display the system's configuration in simulations at packing densities matching experiment. It is evident from the figures that as the packing density increases the system tends to include more clusters of larger sizes, qualitatively aligning with experimental observations. Figures ~\ref{fig_self_assembly}(p)-\ref{fig_self_assembly}(t) illustrate the distribution of cluster size $n_{c}$ (the number of particles in a cluster) in both experiments and simulations. The distributions from both experiments and simulations reveal that with increasing packing density, the number of small clusters decreases, while the number of larger clusters gradually increases. 

We proceed to analyze particle rotation during the self-assembly process. We assign each particle a pointing direction at the beginning (0 min) and track its direction throughout the evolution. Figures ~\ref{fig_rotation_angle}(a)-\ref{fig_rotation_angle}(d) plot the distribution of rotation angle $\theta$ at times as shown in Figs.~\ref{fig_self_assembly}(b)-\ref{fig_self_assembly}(e). Here, we only consider the magnitude of the rotation, but not its direction. As can be seen from the figure, over the evolution, particles with larger rotation angles emerge. However, the rotation remains relatively small. Notably, except for one particle with $\theta$ slightly above 90 degrees in Fig.~\ref{fig_rotation_angle}(a) and reaching 350 to 360 degrees in Figs.~\ref{fig_rotation_angle}(c) and \ref{fig_rotation_angle}(d), all other particles exhibit rotation angles below 90 degrees in Figs.~\ref{fig_rotation_angle}(a)-\ref{fig_rotation_angle}(d). Figures ~\ref{fig_rotation_angle}(e)-\ref{fig_rotation_angle}(h) plot $\theta$ obtained from simulation at the matching densities as in ~\ref{fig_rotation_angle}(a)-\ref{fig_rotation_angle}(d). We can see that simulation results exhibit a similar trend.

We show the statistics of the average cluster size $<n_{c}>$ and the average rotation angle $<\theta>$ as a function of packing density $\phi$ in Fig.~\ref{fig_average_size}. As can be seen from Fig.~\ref{fig_average_size}(a), the initial average number of particles in a cluster is approximately 1.3; the final value in experiments goes up to about 2.4. The average number of particles in a cluster in both experiments and simulations demonstrates qualitative agreement. Regarding the average rotation angle, the average rotation angle from the experiment has a significant jump at packing density $\phi=0.27$ and shows a value of about 17.4 degrees (red circle); after that, the increasing trend decreases, with $<\theta>$ for the final configuration at about 27.6 degrees. The simulation results show a similar trend. If we exclude the particle that has a super-large rotation angle and then calculate the average, the experimental results (green circle) and the simulation results are closer. This observation, that particles have a relatively small average rotation angle, alongside the tendency for clusters to form from nearby particles, suggests cluster growth occurs through the absorption of similarly oriented particles or small clusters, with minimal movement or rotation. When particles or clusters adhere together, they show more similar orientations. However, as packing density increases, misaligned clusters unable to adhere tend to stack, resulting in the final configuration comprising clusters of varied orientations [Fig.~\ref{fig_average_size}(c)]

The Cheerios effect between floating particles involves both interfacial and gravitational energies. The capillary forces are also believed to drive the self-organization in active systems such as bacterial colonies \cite{Black}.  A dimensionless parameter, the Bond number, $B=\rho g L^2/\gamma$, where $L$ is a characteristic length of the particle, $\rho$ is the density of the liquid, $g$ is the gravitational constant, and $\gamma$ is the surface tension, measures the relative importance of gravitational and surface tension effects. Large $B$ corresponds to large particles or a small surface tension coefficient. For an infinite circular cylinder floating on the surface of a liquid, the characteristic length is usually taken as the radius $R$. In our study, we have a square cross-section (7 mm in width and 7 mm in height). Taking the circumradius as an estimation, i.e., $L = 7\sqrt{2}/2$ mm,  the Bond number in our case is approximately $B \approx 7$. This value indicates that buoyancy effects need to be take into consideration to correctly interpret the attractive force \cite{Mahadevan2005}. References \cite{ALLAIN,ALLAIN199326} show that for two infinitely long cylinders floating parallel on the liquid surface, when the Bond number is small ($B \leq 1$), the scaled total interaction energy per unit length, $F/\gamma B^2$, decays exponentially with the scaled distance $l^{*}=l/\lambda$, where $l$ is the center-to-center distance and $\lambda$ is the capillary length, defined as $\lambda = \sqrt{\gamma/\rho g}$. In the range $1 \leq B \leq  10$, the curves show slight deviations from the small number solution. In our case, $\lambda \approx 1.9$ mm. We thus expect that for particles or clusters separated by a distance much greater than the capillary length, the interaction will be very weak. The shrinkage of the surface, however, reduces their separation, enhancing the Cheerios effect and promoting the aggregation of clusters. We also tested particles with varying densities and observed that the behavior of the average cluster size follows a similar trend to the results presented here. Additional details can be found in Appendix \ref{appendix_various_density}.

In our system, once two particles or a particle and a cluster adhere, the bonds remain intact. We rarely observe changes in overlap length, except at relatively high packing densities where particles become more compressed. Collisions between clusters may increase the overlap length. This stable bonding is attributed to the strong interaction upon contact \cite{Mahadevan2005}, which stabilizes the adhered structure and causes it to move as a single entity. The phenomenon of self-assembly of anisotropic particles at fluid interfaces has also been reported for colloidal particles. For instance, Ref.~\cite{PhysRevLett.94.018301} described micron-sized prolate ellipsoids trapped at an oil-water interface. Depending on surface chemistry, these particles aligned preferentially either side-to-side or tip-to-tip. In our system, we also observe some tip-to-tip alignment; however, side-to-side alignment occurs more frequently. 

\vspace{-0.8em} % 减少段间距
\section{\label{sec:level1}Conclusion}
Inspired by the observation of fallen pine needles self-assembling into clusters in small puddles after rainfall, we conducted an experimental investigation of the dynamic evolution of synthetic particles as the liquid surface itself undergoes shrinkage. We observe that particles tend to adhere to one another, forming distinct clusters as the surface area reduces. We identify the emergence of clusters and systematically explore how cluster sizes and the number of clusters evolve as a function of packing fraction. We also analyze particle rotation during the process and discuss the formation of the final configuration comprising clusters of various orientations. Complementary numerical simulations that capture the two main factors, the adhesion between particles and the reduction of surface area, demonstrate qualitative agreement with our experimental findings. Our study contributes to a better understanding of the self-assembly phenomena occurring when macroscopic particles interact on a liquid surface with a dynamically evolving area.

\begin{acknowledgments}
We thank Prof.~Cristina Marchetti for a thorough review of the manuscript and valuable suggestions. This work was supported by the National Natural Science Foundation of China (Grant No.~12274330) and the Knowledge Innovation Program of Wuhan-Shuguang (Grant No.~2022010801020125). D.W. also acknowledges the ``Xiaomi Young Scholar Program" at Wuhan University. This research was supported in part by grant NSF PHY-2309135 to the Kavli Institute for Theoretical Physics (KITP).

\end{acknowledgments}

\section*{DATA AVAILABILITY}
The data that support the findings of this article are openly available  \cite{lx2024}.

\appendix

\section{\label{sec:level1} Structure and floating behavior of a particle}  \label{appendix_particle_structure} 
\begin{figure}[htp]
\centering
\includegraphics[width=.9\linewidth]{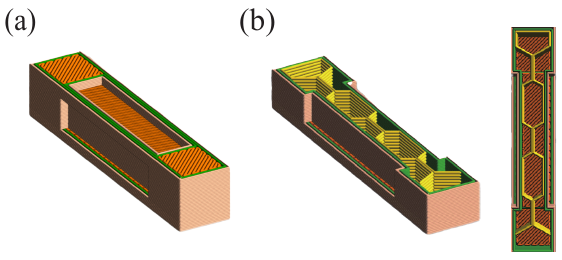}
\caption{Structure of a particle. (a) External structure. Each face of the particle exhibits a central groove. Solid fills are denoted in orange, inner walls in green, and outer walls in flesh color. (b) Internal structure. The particle's hollow interior reveals a hexagonal filling shape (highlighted in yellow). The cross-sectional view is presented on the right. Colors maintain consistent color coding as in (a).}
\label{fig_particle}
\end{figure}

Figure~\ref{fig_particle} illustrates the particle structure. Each particle is 35 mm × 7 mm × 7 mm, with a centrally located groove (20 mm × 4 mm × 1 mm) on each of its four major faces. The particles have a 15\% filling ratio, enhancing their buoyancy. Note that the colors in Fig.~\ref{fig_particle} are for illustration purposes; the actual particles are white, with grooves marked in red for image analysis.  

\begin{figure}[htp]
\centering
\includegraphics[width=.9\linewidth]{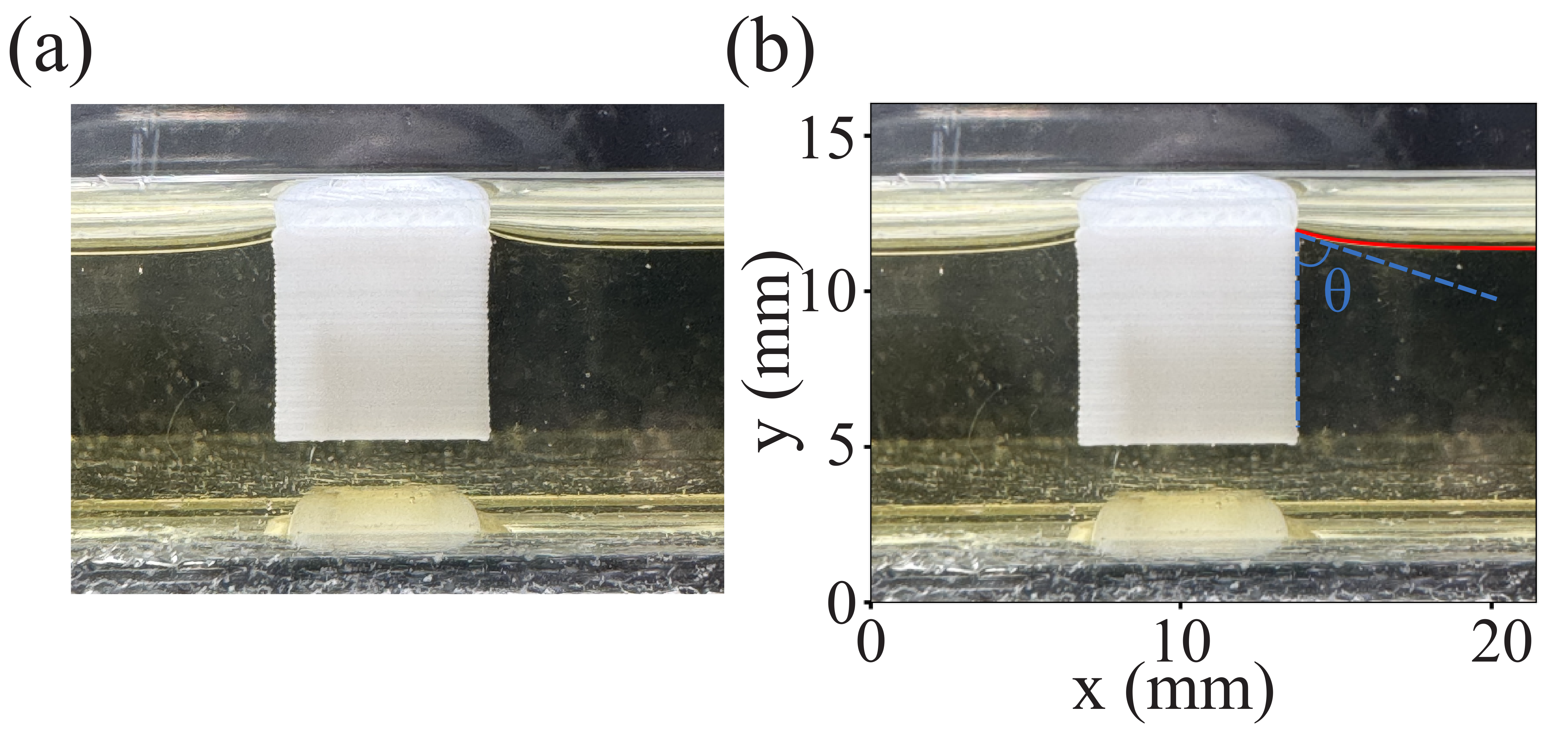}
\caption{Meniscus deformation around a single particle floating on the oil surface. (a) Original photograph of the floating particle. (b) Photograph showing the coordinate system, the estimated contact angle $\theta$, and the fitted curve (red line) representing the surface deformation.}
\label{fig_meniscus}
\end{figure}

To visualize the meniscus deformation around a particle, we prepared a transparent bowl filled with water and topped with a layer of oil. We placed a particle near the wall of the bowl, with its smaller side facing the wall. We then captured an image of the smaller side from outside the bowl wall. Figure ~\ref{fig_meniscus}(a) shows the image of the floating particle. In Fig.~\ref{fig_meniscus}(b), we show the estimated contact angle $\theta$, which was measured to be approximately 72°. The deformation of the surface was fitted using the following equation \cite{Mahadevan2005}:
\begin{equation}
{y = } {\lambda} {\cot(\theta)} \cdot {\exp\left(\frac{-(x-x_0)}{\lambda}\right) + }  {h_0}.
\end{equation}
\noindent { Substituting $x_0 = 13.7$ mm, $h_0 = 11.4$ mm, which were measured from the photograph, and $\theta = 72^\circ$, the fitting yields a capillary length of $\lambda = 1.82$ mm. The fitted curve is shown as the red line in Fig.~\ref{fig_meniscus}(b).}

\section{\label{sec:level1} Two-particle attraction}  \label{appendix_two_particle_attraction}
\begin{figure}[htp]
\centering
\includegraphics[width=.9\linewidth]{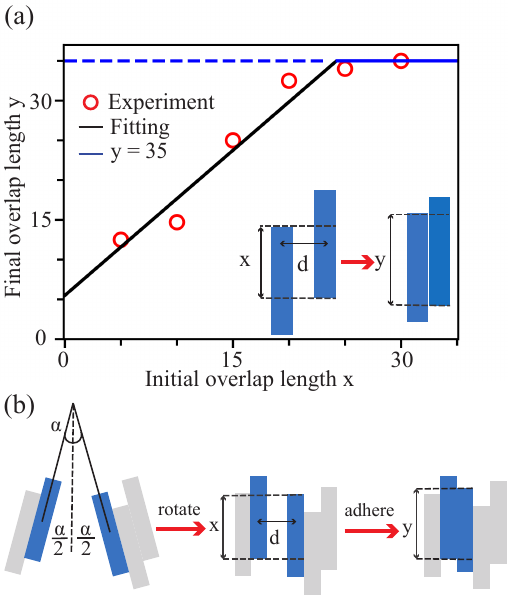}
\caption{Adhesion of two particles. (a) Final overlap length $y$, as a function of initial overlap length $x$, when two particles are released at a fixed horizontal distance $d=1.2$ cm. The inset illustrates the initial and final states. The black line represents the linear fit of the experimental data based on the first five data points, while the blue line denotes the particle length. (b) Determination of the adhesion of two clusters in simulations. When the outermost particles (marked in blue) of two adjacent clusters form an angle $\alpha \leq 5^{\circ}$, we rotate each cluster by $\alpha/2$ to align them in parallel. If the horizontal distance $d \leq 1.2$ cm and the vertical overlap length $x$ is greater than 1/7 of the particle length, the two clusters are allowed to adhere with a probability of $p = 0.02$. The final vertical overlap length $y$ is determined by the black and blue lines in (a).}
\label{fig_adhesion}
\end{figure}

To investigate the attraction between two particles, we align them parallel to each other with their large faces facing each other. Maintaining a horizontal separation of $d = 1.2$ cm, we systematically vary the initial vertical overlap length $x$ from 1/7 to 6/7 of the particle length, in increments of 1/7 as equal divisions. Upon releasing the particles, they adhere to each other, and we measure the final overlap length $y$ for each initial overlap length $x$, as depicted in Fig.~\ref{fig_adhesion}(a). Note that the particles do not necessarily remain parallel during the adhesion process, particularly when they are closer. Often, one end of a particle adheres to the other particle first, followed by the rest of the particle adhering subsequently.
The black line represents the linear fit of the first five experimental data points, where the final overlap $y$ values are less than the particle length. The blue line denotes the particle length.

\section{\label{sec:level1} Simulation details} \label{appendix_simulation_details}
We employ Monte Carlo simulations to investigate the self-assembly of two-dimensional PLA particles as the overall area shrinks. These simulations adhere to the standard hard particle Monte Carlo approach, with the inclusion of an adhesion mechanism to replicate attractive interactions, commonly known as the Cheerios effect \cite{Mahadevan2005, Protiere2023}. The particles are modeled as two-dimensional rectangles, and each Monte Carlo step involves $N$ trial moves, encompassing translations and rotations of the particles \cite{Haji-Akbari2009, Wan2021}. We set the length of the long side of the particle as the unit length in simulations. Translation steps are fixed at 0.063, approximately 2.2 mm, while rotation steps are 0.033 radians, roughly 1.9 degrees. Overlap checks between particles are conducted using the hard particle Monte Carlo module in HOOMD-blue \cite{Anderson2016}, and moves resulting in overlap or boundary intersection are rejected.

\begin{figure}[htp]
\centering
\includegraphics[width=.9\linewidth]{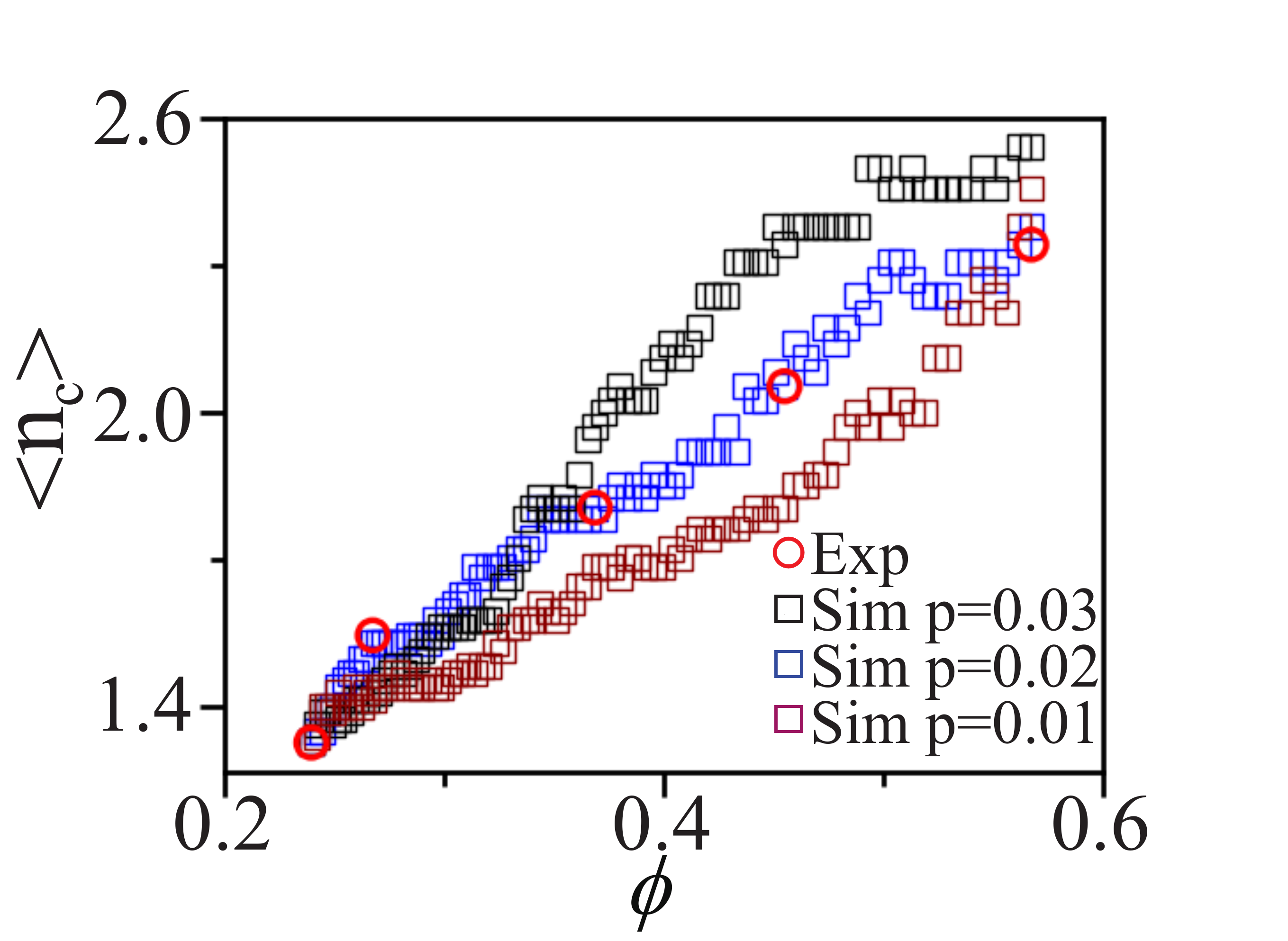}
\caption{The average cluster size $\langle n_c \rangle$ as a function of packaging density $\phi$ for various $p$ values.  }
\label{fig_different_p}
\end{figure}

To account for the attraction between particles, we incorporate an adhesion mechanism. For simplicity, we consider adhesion only between the long sides of particles and attraction between two particles. Specifically, when the outermost particles of two adjacent clusters (or individual particles) form an angle of $\alpha \leq 5^{\circ}$, we rotate each cluster by $\alpha/2$ to align them in parallel [see an illustration in Fig.~\ref{fig_adhesion}(b)]. If the horizontal distance $d \leq 1.2$ cm and the vertical overlap length $x$ exceeds 5 mm (1/7 of the particle length), we allow the two clusters to adhere to each other with a probability of $p = 0.02$. The value of $p$ is chosen to match experimental data. As shown in Fig.~\ref{fig_different_p}, varying $p$ influences the average cluster size. This figure presents simulations starting from the same initial configuration but with different $p$ values, while all other parameters remain unchanged. The final vertical overlap length $y$ is determined by the black line fitting in Fig.~\ref{fig_adhesion}(a) if $y$ is less than the particle length; otherwise, $y$ is set to the particle length, as indicated by the blue line in the figure.

We initiate simulations with $N = 150$ rectangular particles, maintaining the same aspect ratio as experimental conditions, within a circular boundary. To closely mimic the experimental setup, we initialize the system with the same configuration as in the experiment. To accommodate liquid surface shrinkage, the circular boundary is rescaled to 0.995 times its initial value. If the rescaled boundary overlaps with particles, rescaling is rejected, and 10 Monte Carlo steps with the adhesion mechanism are performed. This process is repeated until successful boundary rescaling. Configurations are recorded at specific packing densities.

\section{\label{sec:level1} Identification of clusters}  \label{appendix_identification_clusters}
We identify particles belonging to the same cluster in both experiments and simulations based on the following criteria: (1) The orientation difference between two particles is less than 0.4 radians; (2) The distance between the mass centers of the two particles is less than 0.89 units; (3) The horizontal distance and vertical overlap length after aligning the two particles parallel [denoted as $d$ and $x$ in Fig.~\ref{fig_adhesion}(b)] are less than 0.268 units and greater than 0.37 units, respectively.\\

\section{\label{sec:level1} Experiments with particles of various densities} \label{appendix_various_density}
We placed two particles initially parallel on the corn oil surface in a head-to-head, end-to-end configuration, meaning the initial overlap length $x$ equals the particle length and the horizontal distance $d$ represents the center-to-center separation. We investigated the attraction time—the time required for two particles to fully adhere—using particles with three different densities: $\rho_1 = 0.65~\text{g}/\text{cm}^3$, $\rho_2 = 0.75~\text{g}/\text{cm}^3$, and $\rho_3 = 0.86~\text{g}/\text{cm}^3$. The particles with $\rho_2 = 0.75~\text{g}/\text{cm}^3$ correspond to those described in the main text. Table~\ref{table_fitting_time} summarizes the attraction time for various initial separations $d$. For the particles with $\rho_2 = 0.75 \, \text{g}/\text{cm}^3$, when the initial separation is 2.7 cm, their behavior differs from that observed at smaller initial separations. In this case, their trajectories are easily perturbed by noise after prolonged waiting times, and no reliable trajectory is observed over repeated attempts. Therefore, particles with initial separations of 2.7 cm or greater are not expected to come into contact within the experimental timescale unless additional factors, such as surface area reduction, are considered. When comparing the attraction times of particles with different densities, we observe that lighter particles consistently require less time to come into contact.

\begin{table}[h]
\renewcommand{\thetable}{\arabic{table}}
\captionsetup{labelformat=empty}
\centering
\caption{Table 1. Attraction time (in minutes) for two particles initially separated by a distance $d$.}
\renewcommand{\arraystretch}{1.3} % 控制表格行距
\resizebox{0.95\linewidth}{!}{ % 让表格缩小到 95% 宽度
\begin{tabular}{ c | c | c | c }
 \hline
 \diagbox{Initial distance \\ $d$ (cm)}{Particle density\\ $(\text{g}/\text{cm}^3)$} 
 & $\rho_1 =0.65 $ 
 & \makecell{\raisebox{-1em}{$\rho_2 =0.75$} \\ (main text)} 
 & $\rho_3 =0.86$ \\
 \hline
 1.2 & 1 & 2 & 5 \\
 \hline
 1.7 & 2 & 5 & 10 \\
 \hline
 2.2 & 6 & 24 & 30 \\
 \hline
 2.7 & 21 & No traj. & No traj. \\
 \hline
\end{tabular}
}
\label{table_fitting_time}
\end{table}

\begin{figure}[htp]
\centering
\includegraphics[width=1\linewidth]{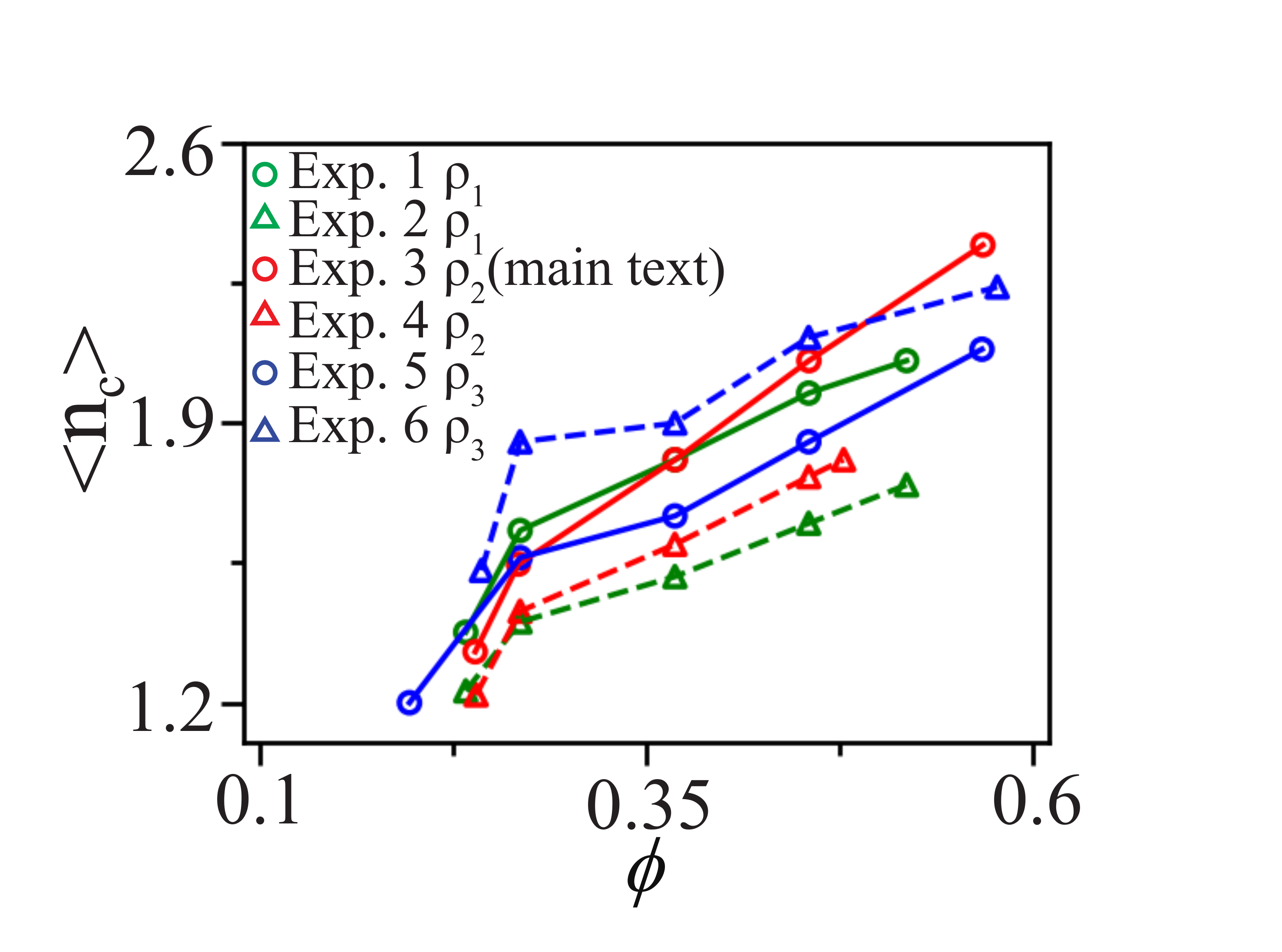}

\caption{The experimental average cluster size $<n_{c}>$ as a function of packing density for three particle densities ($\rho_1 = 0.65\text{g}/\text{cm}^3$, $\rho_2 = 0.75\text{g}/\text{cm}^3$, $\rho_3 = 0.86\text{g}/\text{cm}^3$). The red circles indicate data referenced in the main text.}
\label{fig_different_density}
\end{figure}

Figure ~\ref{fig_different_density} shows clustering experiments for particles of three different densities. The results indicate a similar trend, where the average cluster size increases with packing density. However, due to differences in initial configurations and the limited number of experimental attempts, we cannot draw definitive conclusions about the influence of particle density on clustering behavior.

\bibliography{apssamp}% Produces the bibliography via BibTeX.

\end{document}